\documentclass[12pt,fullpage,fleqn,psfig]{article}
\usepackage{latexsym}
\usepackage{epsfig}
\large
\newcommand{\sumint}{\rlap{\hspace{-.5mm} $\sum$} \int}
\newcommand{\beq}{\begin{equation}}
\newcommand{\eeq}{\end{equation}}
\newcommand{\ba}{\begin{array}{ccc}}
\newcommand{\ea}{\end{array}}
\newcommand{\nn}{\nonumber}
\newcommand{\noi}{\vspace{12pt}\noindent}
\newcommand{\be}{\begin{eqnarray}}
\newcommand{\ee}{\end{eqnarray}}

\newcommand{\oS}{\overline\Sigma}
\begin{document}
\newcommand{\mat}{\left ( \begin{array}{cc}}
\newcommand{\emat}{\end{array} \right )}
\newcommand{\matt}{\left ( \begin{array}{ccc}}
\newcommand{\ematt}{\end{array} \right )}
\newcommand{\matf}{\left ( \begin{array}{cccc}}
\newcommand{\ematf}{\end{array} right )}
\newcommand{\vect}{\left ( \begin{array}{c}}
\newcommand{\evect}{\end{array} \right )}
\newcommand{\Tr} {\rm Tr}
\newcommand{\cotanh}{{\rm cotanh}}
\def\beqn{\begin{eqnarray}}
\def\eeqn{\end{eqnarray}}
\def\la{\lambda}
\def\ga{\gamma}
\def\om{\omega}
\def\al{\alpha}
\def\d{\partial}
\def\Tr{ {\rm Tr} }

\renewcommand{\baselinestretch}{2}

\thispagestyle{empty}
\parskip=4mm
\newcommand{\R}{{\mathchoice{\hbox{$\sf\textstyle I\hspace{-.15em}R$}}
{\hbox{$\sf\textstyle I\hspace{-.15em}R$}} {\hbox{$\sf\scriptstyle
I\hspace{-.10em}R$}} {\hbox{$\sf\scriptscriptstyle
I\hspace{-.11em}R$}}}}

\renewcommand{\theequation}{\arabic{section}.\arabic{equation}}

\hfill{SUNY-NTG-02/15}

\vspace{1cm}
\begin{center} 
{\Large\bf Thermodynamics of chiral symmetry at low densities}
\vspace{8mm}

K. Splittorff$^1$, D. Toublan$^2$ and J.J.M. Verbaarschot$^3$

{\small
$^1$The Niels Bohr Institute, Blegdamsvej 17, Copenhagen, Denmark \\
$^2$Physics Department, University of Illinois at Urbana-Champaign,
Urbana, IL 61801, USA\\  
$^3$Department of Physics and Astronomy, SUNY, Stony Brook,
NY\,11794, USA\\} 
\vskip 3cm

{\bf Abstract} \\

\end{center}

\noi
The phase diagram of two-color QCD as a function of temperature and baryon 
chemical potential is considered. Using a low-energy chiral Lagrangian
based on the symmetries of the microscopic theory, 
we determine, at the one-loop level, the temperature dependence of the critical
chemical potential for diquark condensation and the temperature dependence 
of the diquark condensate and baryon density. The prediction for the
temperature dependence of the critical chemical potential is consistent with
the one obtained for a dilute Bose gas. 
The associated phase transition is
shown to be of second order for low temperatures and first order at higher 
temperatures. The tricritical point at which the second order phase 
transition ends is determined. The results are carried over to QCD with 
quarks in the adjoint representation and to ordinary QCD at a non-zero
chemical  potential for isospin.

\vfill


\setcounter{page}{0}
\newpage
\setcounter{equation}{0}
\section{Introduction}

The properties of strongly interacting matter at finite density and
temperature are presently being studied 
in a variety of ways. Phenomenologically, our understanding of this subject 
is based on heavy ion collisions and the properties of neutron stars.
On the theoretical side several approaches are followed. An asymptotic
expansion in the inverse baryon density can be derived in the framework of 
perturbative QCD \cite{Bailin,Son}. At realistic densities, first principle
calculations are not possible and one has to rely on schematic models
such as the Nambu-Jona-Lasinio model \cite{Krishna,CG,ABR},
Random Matrix Models \cite{misha,tricrit},
or semiclassical methods such as the instanton liquid model \cite{RSSV}.
Based on these studies a plethora of phases and phenomena
have been proposed for QCD at non-zero baryon density. Among others, we 
mention a superconducting phase \cite{Bailin,RSSV,Krishna}, 
a color-flavor locked
phase \cite{ARW}, a crystalline phase \cite{ABR}, 
a tricritical point \cite{tricrit,krish3}
and continuity between  quark and hadronic matter 
\cite{Thomas}.

Because of the complex phase of the fermion determinant these
scenarios cannot be  
confirmed by means of first principle lattice simulations of QCD.
This makes it imperative to analyze closely related theories for
which lattice simulations at realistic densities are possible. We will consider
QCD with two colors at  non-zero baryon density, QCD with
three colors at non-zero isospin density
(or phase quenched QCD at non-zero baryon density) or, 
in general, at nonzero flavor density, and QCD with gauge fields in the adjoint
representation at nonzero baryon density.
Chiral symmetry is broken spontaneously in the ground state of both
QCD with two colors and QCD with three colors, and the low-energy
excitations of these theories are given by a theory of weakly 
interacting Goldstone
bosons which can be analyzed order by order in Chiral Perturbation Theory
\cite{WeinbergEff,GaL}. Within this framework, a phase transition 
to a Bose condensed 
phase has been found to be induced by a chemical potential related to
a  quantum number carried by at least one  of the Goldstone bosons
\cite{KST,KSTVZ,TV,Misha-Son,SSS,Dominique}.  

For QCD with three colors 
this corresponds to the inclusion of a chemical
potential for isospin or strangeness leading to pion or kaon condensation
on a scale determined by the mass the Goldstone modes
\cite{Misha-Son,Dominique}. Because of the Pauli-G\"ursey symmetry,
Goldstone bosons with non-zero baryon number appear in QCD with two
colors, leading to diquark condensation for a baryon chemical potential
beyond half the mass of the diquark boson \cite{KST,KSTVZ,RSSV,STV1}. By
including an isospin chemical potential the combined phase diagram
can be studied within the framework of  chiral perturbation theory
\cite{Misha-Son,SSS}. We emphasize that in the chiral limit these
results are exact and should coincide with the microscopic theory. 

In the color-flavor locked phase which has been
proposed as a possible state of QCD at high density
chiral symmetry is broken spontaneously as well. The corresponding
Goldstone bosons can be described in terms of a chiral Lagrangian 
\cite{ARW,CG,SS-CFL,BedSav} very much like the one for 
QCD with two colors 
\cite{KST,KSTVZ}.

In this article we study chiral Lagrangians at non-zero chemical potential
and temperature. We mainly  focus on QCD with two-colors for
which also lattice QCD simulations are 
available \cite{SU2lattice.old,SU2lattice.new,LATTICE}, but we will discuss
the extension of our results to QCD with adjoint quarks at finite baryon
density and to QCD with three or more colors and fundamental quarks at
non-zero isospin density.
In a previous publication \cite{STV1} we have shown that the chiral Lagrangian 
for QCD with two colors is renormalizable to one loop, i.e., all infinities
that are generated by the one-loop diagrams can be absorbed into
the coupling constants of the next to leading order terms in the chiral
Lagrangian such that the constants are independent of the temperature
and the chemical potential. In that article  we have also shown that,
to one-loop order, the critical chemical potential is given by half the {\em
renormalized} pion mass.

Recently Kogut and Sinclair \cite{KS1} have investigated the critical 
exponents of the phase transition at zero temperature in quenched QCD with
three colors at finite isospin chemical potential, $\mu_I$, 
and in QCD with two
colors at finite baryon chemical potential, $\mu_B$. From chiral perturbation
theory at one-loop level we expect that at zero temperature the phase
transition to the charged phase is characterized by mean field critical 
exponents. The measurements in \cite{KS1} are consistent with mean field
critical exponents.  
Lattice simulations \cite{KTS} at non-zero baryon chemical potential 
and temperature, $T$,
indicate that QCD with two colors 
displays a rich and interesting phase diagram including
a tricritical point in the $(\mu_B,T)$-plane.
In this article we explore this phase diagram by means of chiral perturbation
theory by extending the 1-loop calculation in \cite{STV1} to non-zero
temperature. Our results support the existence of a
tricritical point in the ($\mu_B,T$)-plane.

The phase diagram of three-color QCD in the ($\mu_I,T$)-plane has
been studied recently on the lattice \cite{KS2}. The phase diagram again
includes a tricritical point. Extending the analytic results for $N_c=2$
to $N_c=3$ we confirm the existence of this tricritical point in the
($\mu_I,T$)-plane.


The pattern of chiral symmetry breaking is determined by the Dyson index
$\beta_D$ of the Dirac operator. Its value denotes the number of independent
degrees of freedom per matrix element. 
If the Dirac operator 
does not have anti-unitary symmetries its matrix elements are complex
and $\beta_D =2$. This is the case
for QCD with three or more colors and fundamental fermions. 
Anti-unitary symmetry operators can be
written as $AK$ with $A$ a unitary operator and $K$
the complex conjugation operator. There are only two nontrivial possibilities
for $(AK)^2$,
\be
(AK)^2 = 1 & {\rm with}  & \beta_D = 1 \ , \\
(AK)^2 = -1 & {\rm with} &  \beta_D = 4 \ .
\ee
In the first case it is always possible to find a basis for which the Dirac
operator is real. This case is realized for QCD with two colors and quarks
in the fundamental representation. The second case, 
in which the matrix elements
of the Dirac operator can be organized into selfdual quaternions, is
realized for QCD with two or more colors and quarks in the adjoint 
representation. The addition of a chemical potential does affect
this triality \cite{mutree}. Below
the critical value of the chemical potential, the number of Goldstone
bosons is determined by the pattern of chiral symmetry breaking at
zero chemical potential. The only thing that happens at nonzero chemical 
potential
is that  degeneracies in  the mass spectrum 
of the Goldstone bosons are lifted. The three different cases are very
similar. The only difference is in the degeneracies of the Goldstone
modes which only depend on the symmetry breaking pattern. For
definiteness we analyze $\beta_D =1 $  in this article, and 
in sections \ref{sec:adjQCD} and \ref{sec:QCDmuI}, we discuss
the changes that occur for $\beta_D =4$ and $\beta_D=2$, respectively.

The paper is organized as follows. To make the paper self-contained we first 
discuss QCD with two colors and the transition into the diquark condensed 
phase at zero temperature. Then we calculate the temperature dependence of 
the free energy in the two phases of the theory separately. Using this we 
determine the critical chemical potential for diquark condensation at 
non-zero temperature and the tricritical point at which the second order
phase transition turns first order. We also study the order
parameter and the number density.

\section{The Symmetries of QCD with Two Colors}
\label{sect:symQC2D}

The Lagrangian that describes the low-energy limit of QCD with two colors is
uniquely determined by the symmetries of the microscopic theory 
and the assumption of spontaneous breaking of chiral symmetry. We
start by discussing the symmetries.

QCD with two colors and fermions in the fundamental representation is
special since the gauge group is pseudo real. 
Its generators, $\tau_k$, satisfy
$-\tau_2\tau_k^\dagger\tau_2=\tau_k$. 
This implies that the Dirac operator, $D$, 
satisfies the anti-unitary symmetry,
\be
D\tau_2C\gamma_5 = \tau_2C\gamma_5D^* \ ,
\ee
where $C$ is the charge conjugation matrix. This symmetry 
follows immediately from the explicit form of the Dirac operator,
\be
D= \gamma_\nu D_\nu +m +\mu_B\gamma_0 \ .
\ee 
Here, $D_\nu \equiv \partial_\nu +A_\nu$ is the gauge covariant derivative 
and $\gamma_\nu$ are the hermitian Euclidean Dirac matrices.
The microscopic Lagrangian at $m=\mu_B=0$ has an $SU(2N_f)$ chiral
invariance,  often referred to as the Pauli-G\"ursey symmetry
\cite{pauli-gursey}. This flavor symmetry is manifest if we choose a
basis given by left-handed 
quarks, $q_L$, and conjugate right-handed anti-quarks, $\sigma_2\tau_2
q_R^*$, ($\sigma_2$ acts in spin space),
\be
\Psi \equiv \left(\begin{array}{c} q_L \\ \sigma_2\tau_2 q_R^*
\end{array}\right) \ . 
\label{basis}
\ee
The fermionic part of the Lagrangian for QCD with two colors at $m=\mu_B=0$ 
in this basis is 
\be
\label{LqcdFree}
L_{\rm QCD} & = & i\Psi^\dagger \sigma_\nu D_\nu\Psi \ ,
\ee
where $\sigma_\nu=(-i,\sigma_k)$. It is invariant under global
transformations 
\be
\Psi \to V\Psi, \; {\rm with} \; V \in SU(2 N_f) \ .
\label{Psirot}
\ee
The global $SU(2 N_f)$ chiral invariance is broken explicitly at nonzero 
baryon chemical potential, $\mu_B$, quark mass, $m$, or the diquark source,
$j$, which can be seen from the QCD Lagrangian in
terms of the $\Psi$-fields, 
\be
\label{Lqcd}
L_{\rm QCD} & = & i\Psi^\dagger \sigma_\nu(D_\nu-\mu_B B \delta_{0\nu})\Psi
-(\frac{1}{2}\Psi^T \sigma_2\tau_2 {\cal M} \Psi + {\rm h.c.}) \ .
\ee
The matrix that includes the mass and the diquark source is defined by 
\cite{KSTVZ},
\be
\label{chi}
  {\cal M}\equiv     \sqrt{m^2+j^2} 
  ({\hat M}  \cos  \phi+ {\hat J} \sin \phi) \ ,
\ee
where $\tan\phi = j/m$ .
The baryon charge matrix $B$, the mass matrix $\hat{M}$, and the diquark
source matrix $\hat J$ are given by,
\be
 B  \equiv \left( \begin{array}{cc} 1&0\\ 0&-1
    \end{array} \right) \ , \ \ {\hat M}  \equiv  
  \left( \begin{array}{cc} 0&1\\-1&0
    \end{array} \right) \  , \ \ \ 
      {\rm and} \ \ \ \hat J\equiv\left(
    \begin{array}{cc} 
      iI&0\\0& iI \end{array} \right) 
\label{MandB}
\ee
with 
\be
I\equiv \left( \begin{array}{cc} 0&-1\\1&0
    \end{array} \right) \ .
\ee
The explicit breaking of the global $SU(2N_f)$ symmetry of the Lagrangian
can be compensated by a rotation of the source matrices,
\begin{eqnarray}
\label{globalQCDtrans}
 {\cal M} &
  \rightarrow  & V^* {\cal M} V^\dagger \ , \\ 
 B & \rightarrow & V B V^\dagger \ .
\end{eqnarray}
The combined transformation of $\Psi, {\cal M}$, and $B$ leaves
$L_{\rm QCD}$ invariant. This global $SU(2N_f)$ invariance of the Lagrangian
can be   extended to a local $SU(2N_f)$ flavor symmetry by introducing the
vector field $B_\nu\equiv(B, 0,0,0)$ with transformation properties
\cite{Alvarez,KST,KSTVZ}, 
\begin{eqnarray}
\label{localQCDtrans}
   B_\nu \rightarrow V B_\nu V^\dagger -\frac1{\mu_B} V \partial_\nu
   V^\dagger \ . 
\end{eqnarray}
This invariance is vital as we impose that the chiral
Lagrangian  must be invariant under this local flavor symmetry.

\section{The chiral Lagrangian}

The global $SU(2N_f)$ chiral symmetry of the microscopic theory is
believed to be spontaneously broken down to $Sp(2N_f)$ by the
vacuum \cite{SmilgaV,TV14}. Associated with this 
spontaneous breakdown is a set of $N_f(2N_f-1)-1$ Goldstone modes described
by the field $\Sigma \in SU(2N_f)/Sp(2N_f)$.
For simplicity, we will only consider an 
even number of degenerate flavors.   
For a discussion of non-degenerate flavors see \cite{Dominique}. 

\noi
Under the local $SU(2N_f)$ rotation the Goldstone field
$\Sigma$ transforms as
\be
\Sigma \to V \Sigma V^T, \qquad V \in SU(2N_f) \ .
\label{Sigmarot}
\ee
The underlying microscopic theory  is invariant under the combined
local transformation (\ref{Psirot}),
(\ref{globalQCDtrans}), (\ref{localQCDtrans}) of $\Psi$, $B$, and ${\cal
  M}$.   The same symmetries should be present
in the low-energy chiral Lagrangian and is thus 
invariant under the combined transformations (\ref{Sigmarot}),
(\ref{globalQCDtrans}), (\ref{localQCDtrans}) of $\Sigma$, $B$, and ${\cal
  M}$. Therefore only the covariant derivative \cite{KST}
\begin{eqnarray}
  \nabla_\nu \Sigma&=&\partial_\nu \Sigma-\mu_B (B_\nu \Sigma+\Sigma 
  B_\nu^T) \ , \nonumber \\
  \nabla_\nu \Sigma^\dagger&=&\partial_\nu \Sigma^\dagger+\mu_B (B_\nu^T
  \Sigma^\dagger+\Sigma^\dagger B_\nu),
\label{CovDeriva}
\end{eqnarray}
and products of ${\cal M}$ and $\Sigma$ can enter in
the chiral Lagrangian. 
The terms that can enter in the chiral Lagrangian are uniquely
determined by this symmetry requirement and Lorentz invariance. 
However, as the number of allowed
terms is infinite one needs to introduce a counting scheme which defines the 
expansion parameters of the theory. In this paper we extend the usual power
counting of chiral perturbation theory giving the same weight to the square
root of the quark mass, the square root of the diquark source, the chemical
potential, and the momenta: 
\be
{\cal O}(p)\sim {\cal O}(\mu_B)\sim {\cal O}(\sqrt{m})
\sim {\cal O}(\sqrt{j} )\sim \frac{1}{L}  \ .
\ee
Within this counting scheme we obtain a well defined expansion. 

Before describing this expansion it is instructive to take a closer look at 
the nature of the Goldstone manifold $SU(2N_f)/Sp(2N_f)$. This manifold has 
$N_f(2N_f-1)-1$ degrees of freedom and can be represented by anti-symmetric
special unitary matrices of size $2N_f \times 2N_f$. 
A convenient parameterization is given by,
\begin{eqnarray}
  \Sigma=U \oS U^T \ ,
\label{Sigma}
\end{eqnarray}
where,
\begin{eqnarray}
  U=\exp (i \Pi/2 F) \hspace{1cm} {\rm and} \hspace{.5cm} \Pi=\pi_a 
  X_a/\sqrt{2 N_f} \ .
\end{eqnarray}
Here, $F$ is the pion decay constant,
the fields $\pi_a$ are the Goldstone modes and the $X_a$ are the
$2 N_f^2-N_f-1$ generators of the coset $SU(2N_f)/Sp(2N_f)$.
They obey the relation
$X_a \oS
=\oS X_a^T$ and are 
normalized according to ${\rm Tr} X_a X_b=2 N_f \delta_{ab}$.
The constant antisymmetric special unitary matrix $\oS$ is the
ground state about which the $\Pi$-fields fluctuate.

\subsection{Leading Order Chiral Lagrangian}

The chiral Lagrangian to leading order, $p^2$, in the momentum
expansion with the  required invariance properties is the non-linear
$\sigma$-model \cite{KSTVZ}, 
\begin{eqnarray}
  \label{L2}
  {\cal L}^{(2)}=\frac{F^2}{2} {\rm Tr} \Big[ \nabla_\nu \Sigma
  \nabla_\nu \Sigma^\dagger - \chi^\dagger \Sigma - \chi
  \Sigma^\dagger \Big] \ ,
\end{eqnarray}
where we have introduced the source term 
\be
\chi = \frac G{F^2} {\cal M}^\dagger \ .
\ee
The constant $G$ is the chiral condensate in the chiral limit. It is tied
to the leading order pion decay constant $F$ and pion mass $M$ through the
Gell-Mann--Oakes--Renner relation, $M^2=m G/F^2$. 
The action $S_2= \int dx {\cal L}^{(2)}$ at order $p^2$, given by the tree
diagrams of ${\cal L}^{(2)}$, was studied in \cite{KST,KSTVZ}.


\subsection{Next-to-Leading Order Chiral Lagrangian}

At next to leading order the total action is given by four terms \cite{STV1},
\begin{eqnarray}
  \label{NLOpartfct}
  S=S_2+S_{1\rm{-loop}}+S_4+S_A \ .
\end{eqnarray}
The contribution to the free energy from the  one-loop diagrams of
${\cal L}^{(2)}$ is denoted $S_{1\rm{-loop}}$, $S_4=\int dx {\cal L}^{(4)}$,
and $S_A$ is the Wess-Zumino functional. 
The part of the chiral Lagrangian that contributes to the free energy 
at order $p^4$ is given by \cite{STV1},
\begin{eqnarray}
  \label{L4}
  {\cal L}^{(4)}&=&-L_0 {\rm Tr} \Big[ \nabla_\nu \Sigma \nabla_\tau
  \Sigma^\dagger  \nabla_\nu \Sigma \nabla_\tau
  \Sigma^\dagger \Big] - L_1 \Big({\rm Tr} \Big[  \nabla_\nu \Sigma
\nabla_\nu  \Sigma^\dagger   \Big]\Big)^2 \nonumber \\
&&- L_2 {\rm Tr} \Big[  \nabla_\nu \Sigma \nabla_\tau
  \Sigma^\dagger   \Big] {\rm Tr} \Big[  \nabla_\nu \Sigma \nabla_\tau
  \Sigma^\dagger   \Big] -L_3 {\rm Tr} \Big[  \Big(\nabla_\nu \Sigma
 \nabla_\nu  \Sigma^\dagger \Big)^2  \Big] \\
&&+L_4 {\rm Tr} \Big[ \chi \Sigma^\dagger
  + \Sigma \chi^\dagger \Big] {\rm Tr} \Big[  \nabla_\nu \Sigma
  \nabla_\nu 
  \Sigma^\dagger   \Big]+ L_5 {\rm Tr} \Big[ \Big(\chi \Sigma^\dagger
  + \Sigma \chi^\dagger \Big) \Big( \nabla_\nu \Sigma \nabla_\nu
  \Sigma^\dagger   \Big)  \Big]  \nn \\
&&- L_6 \Big({\rm Tr} \Big[ \chi \Sigma^\dagger
  + \Sigma \chi^\dagger \Big]\Big)^2-L_7 \Big( {\rm Tr} \Big[
  \chi^\dagger 
  \Sigma-\chi \Sigma^\dagger \Big] \Big)^2
  \nonumber \\
&&-L_8 {\rm Tr} \Big[ \chi \Sigma^\dagger
  \chi \Sigma^\dagger+\Sigma \chi^\dagger \Sigma \chi^\dagger
  \Big]-H_2 {\rm Tr} \Big[ \chi \chi^\dagger  \Big] \ . \nonumber
\end{eqnarray}
The dimensionless constants $L_k$ and $H_2$ include counter terms that cancel
the divergences that appear at one-loop level
(see \cite{STV1} for a detailed calculation). 
The contribution from the
Wess-Zumino functional is independent of temperature to the 1-loop order 
and will be ignored in this article.
For a discussion of the Wess-Zumino term at 
$\mu_B\neq 0$  we refer to \cite{LSS}. 


\section{Zero Temperature}

We now review the predictions obtained from next-to-leading order
chiral perturbation theory 
at zero temperature \cite{STV1}. These results are the basis for the
present work. As usual in 1-loop calculations, a non-zero temperature 
does not introduce additional divergences in the action. Once we have
renormalized the theory to 1-loop order at $T=0$, it remains finite at
nonzero temperature and chemical potential.
In the present paper we study the $T\neq 0$ effects at  order
$p^4$. We 
first review the order $p^2$ results and the corrections at zero
temperature at order
$p^4$.

At  $T=0$ the orientation of the vacuum  to order $p^2$ is given by, 
\begin{eqnarray}
  \label{oS}
  \oS(\alpha)= \Sigma_c\cos \alpha+  \Sigma_d \sin \alpha \ ,
\end{eqnarray}
where,
\begin{eqnarray}
\begin{array}{lll}
\alpha=0 & {\rm if} & \mu_B< M/2 \ , \\
\cos\al=\frac{M^2}{4\mu^2} & {\rm if} & \mu_B> M/2 \ ,
\end{array}
\end{eqnarray}
and,
\begin{eqnarray}
  \Sigma_c\equiv I \ \ \  {\rm and}
\ \ \ \Sigma_d\equiv \left( \begin{array}{cc} iI & 0 \\ 0 & iI
  \end{array} \right) \ . 
\end{eqnarray}
A non-zero value of $\al$ corresponds to diquark condensation
\cite{KSTVZ}. The diquarks are color neutral and the condensed phase is a
superfluid. We refer to the phases as the normal phase ($\al=0$) and the
diquark- or superfluid phase ($\al\neq0$).
At non-zero $j$, the diquark condensate is always non-vanishing.
In this case, the orientation of the vacuum is still given by
$\oS=\Sigma_c\cos \alpha + \Sigma_d\sin \alpha$,
but $\alpha$ is determined by the saddle point equation,
\begin{eqnarray}
  \label{SadPt}
  4 \mu_B^2 \cos \alpha \sin \alpha=\tilde M^2 \sin (\alpha-\phi)  \ ,
\end{eqnarray}
where $\tilde M$ is the leading order pion mass at diquark source 
$j\neq0$,
\be
 \tilde M^2 = \frac{G\sqrt{m^2+j^2}}{F^2} \ .
\ee

To order $p^4$ calculations we need the propagators of ${\cal L}^{(2)}$. 
In the normal phase we have $N_f^2 -1$ modes with zero baryon charge,
which we will call $P$-modes and $N_f(N_f-1)$ diquark modes, which
we will call $Q$ modes.  
Their inverse propagators  are given by \cite{KSTVZ,STV1}, 
\begin{eqnarray}
  \label{invPropN}
  D^P&=&p^2+M^2 \ , 
\end{eqnarray}
and,
\begin{eqnarray}
  D^Q&=&\left( \begin{array}{cc} p^2+M^2-4 \mu_B^2 & 4 i \mu_B p_0 \\ 
                4 i \mu_B p_0 & p^2+M^2-4 \mu_B^2 
\end{array} \right) \ .
\end{eqnarray}
In the phase of condensed diquarks 
four different types of modes can be distinguished:
$N_f (N_f+1)/2$ $P_S$-modes, $(N_f^2-N_f-2)/2$
$P_A$-modes and $N_f (N_f-1)$ complex $Q$-modes. 
The inverse propagators read, 
\begin{eqnarray}
  \label{invPropD}
  D^{P_S}&=&p^2+M_1^2 +\frac 14 M_3^2  \ ,
 \nonumber \\
D^{P_A}&=&p^2+M_2^2 +\frac 14 M_3^2 \ ,
\\
  D^Q&=&\left( \begin{array}{cc} p^2+M_1^2 & i M_3 p_0 \\ 
                i M_3 p_0 & p^2+M_2^2
\end{array} \right) \ , \nonumber
\end{eqnarray}
where we have introduced,
\begin{eqnarray}
\label{massQ}
M_1^2&\equiv&\tilde M^2 \cos (\alpha-\phi)-4 \mu_B^2 \cos 2\alpha \ , \nonumber
\\ 
M_2^2&\equiv &\tilde M^2 \cos (\alpha-\phi)-4 \mu_B^2 \cos^2\alpha \ , \\
M_3^2&\equiv &16 \mu_B^2 \cos^2 \alpha \ . \nonumber
\end{eqnarray}
The $Q$-modes are mixed, but the mixing matrix can  be
diagonalized analytically with eigenmodes denoted by  
$\tilde Q$ and $\tilde Q^\dagger$. The $\tilde Q$-modes
are of special interest since they are the
Goldstone modes of the superfluid phase at $j=0$ and
$\mu_B>M/2$. In particular, these diquark modes are massless even though
$m\neq0$.

In \cite{STV1}, it was found that at $T=0$ the $p^4$ action is
renormalizable at the 1-loop level. The renormalized $p^4$ action predicts
that for  $\mu_B\geq\mu_c=m_\pi/2$ the diquark condensate is nonzero.
The quantity $m_\pi$ is the physical 
pion mass at 1-loop order expressed in terms of the renormalized couplings of 
${\cal L}^{(4)}$, the ultraviolet cutoff $\Lambda$, and $M^2=Gm/F^2$ 
\cite{STV1},
\be 
m_\pi^2=M^2 \left( 1+\frac{M^2}{F^2} 
 \left[ 4(-2N_f L^r_4- L^r_5+4N_f L^r_6+2 L^r_8)+\frac{N_f+1}
   {64  \pi^2 N_f}  \ln \frac{M^2}{\Lambda^2} \right] \right) \ .
\ee 
The difference between $M^2$ and $m_\pi^2$ is of order $p^4$. Replacing 
$M^2$ with $m_\pi^2$ in the calculation of $S_{\rm 1-loop}$ and $S_4$
therefore only leads to a higher order correction. We shall make this 
replacement throughout the calculation.

\section{Non-Zero Temperature}

The temperature dependence in the chiral expansion enters through $S_{\rm
  1-loop}$ since this is the only part which involves an explicit momentum
integration. In the following sections we calculate the temperature dependence 
of $S_{\rm 1-loop}$ in the normal phase and in the diquark phase, in a
similar way as for usual Chiral Perturbation Theory
at nonzero temperature and zero chemical potential \cite{TempChPT}. This
subsequently allows us determine the critical chemical potential, $\mu_c(T)$,
which separates the two phases.

\subsection{The Normal Phase at $T\neq 0$}

The free energy at 1-loop order in the normal phase is given by
\begin{eqnarray}
  \label{FrEnN}
  \Omega&=& -2  N_f F^2 M^2 - \frac12 (N_f^2-1) G_0^P(m_\pi,T)
-\frac14 N_f (N_f-1) G_0^Q(m_\pi,\mu_B,T) \\
&&-2 N_f m_\pi^4  \Big( 8 N_f L_6+2 L_8 +H_2 \Big) + {\cal O}(p^6) \ ,
\nonumber  
\end{eqnarray}
with 
\be
G_0^P(m_\pi,T) & = & - \sumint  \frac{d^dp}{(2 \pi)^d} \ln (p^2+m_\pi^2), \\
G_0^Q(m_\pi,\mu_B,T) & = & - \sumint  \frac{d^dp}{(2 \pi)^d} \ln \Big(
(p^2+m_\pi^2-4   \mu_B^2)^2 
     +16 \mu_B^2 p_0^2 \Big)  \ .
\ee
In the expressions above we have introduced the notation
\beq
\sumint  \frac{d^dp}{(2 \pi)^d} f(p_0,\vec{p}) \equiv
T\sum_{n=-\infty}^{\infty} \int  \frac{d^{p-1}p}{(2 \pi)^{d-1}} f(2\pi n
T,\vec{p}) \ . 
\eeq
As shown below, the 1-loop contribution can be decomposed in the following way:
\be
G_0(T) & = & \Delta_0+g_0(T) \ ,
\ee
where $\Delta_0$ is the zero temperature propagator at the origin, and
$g_0(T)$ contains the full temperature dependence. The same decomposition
holds in the diquark condensation phase and we shall employ it
throughout this paper.  
The propagator of the $P$-mode at finite temperature can be rewritten
as, 
\begin{eqnarray}
G_0^P(m_\pi,T)&=& \int_0^\infty dt \frac{e^{-t m_\pi^2}}{(4 \pi)^{d/2} t^{d/2+1}} 
\sum_{n=-\infty}^{\infty} e^{-n^2/4 t T^2}.
\label{PN}
\end{eqnarray}
The $T$-independent part is given by the $n=0$ term and can be written as,
\begin{eqnarray}
  \Delta_0^P&=&-\int \frac{d^dp}{(2 \pi)^d} \ln (p^2+m_\pi^2) 
=\frac1{(4 \pi)^{d/2}} \Gamma(-\frac d2) m_\pi^d \ .
\label{DelPN}
\end{eqnarray}
This divergence is canceled by the counter terms in ${\cal L}^{(4)}$
(\ref{L4}) as was explicitly shown in \cite{STV1}.
After performing the $t$-integration over the $n\neq 0$ terms we find
for the temperature dependent part, 
\begin{eqnarray}
  g_0^P(T) &=& 16\frac{m_\pi^2T^2}{(4\pi)^2}\sum_{n=1}^\infty 
\frac{K_2(\frac{m_\pi n}{T})}{n^2} \ ,
\label{gPN}
\end{eqnarray}
where $K_2$ is the modified Bessel function. 
To evaluate $G_0^Q$ it is helpful to use the identity,
\begin{eqnarray}
 (p^2+m_\pi^2-4 \mu_B^2)^2 +16 \mu_B^2 p_0^2&& \\
 &&\hspace{-4cm} = \Big( (p_0-2 i\mu_B)^2+\vec{p}^2+m_\pi^2 \Big)
 \Big((p_0+2 i\mu_B)^2  +\vec{p}^2+m_\pi^2 \Big) \ . \nn 
\label{helpfulEq}
\end{eqnarray}
This results in,
\begin{eqnarray}
  G_0^Q=\frac2{(4 \pi)^{d/2}} \int_0^\infty dt \frac{ 
             e^{-t m_\pi^2} }{t^{1+d/2}}
           \sum_{n=-\infty}^\infty \cosh(\frac{2 \mu_B n}{T}) 
  e^{-n^2/4 t T^2}.
\label{QN}
\end{eqnarray}
The $T$-independent part of this function diverges in 4 dimensions,
\begin{eqnarray}
  \Delta_0^Q&=&-\int \frac{d^dp}{(2 \pi)^d} \ln \Big( (p^2+m_\pi^2-4
  \mu_B^2)^2 
     +16 \mu_B^2 p_0^2 \Big)
=\frac2{(4 \pi)^{d/2}} \Gamma(-\frac d2) m_\pi^d \ ,
\end{eqnarray}
and is renormalized by counter terms in ${\cal L}^{(4)}$.
The temperature dependent part reads,
\begin{eqnarray}
g_0^Q(T)&=& 32 \frac{m_\pi^2T^2}{(4\pi)^2}\sum_{n=1}^{\infty} \frac{{\rm
  K}_2(\frac{m_\pi n}{T})}{n^2} \cosh(\frac{2\mu_B n}{T}) \ .
\label{gQ3} 
\end{eqnarray}
The temperature dependent parts $g_0^P(T)$ and $g_0^Q(T)$ are finite
but are not known  analytically for arbitrary  mass and chemical
potential. However,  two interesting limits can be analyzed analytically:
$T \ll \mu_B, m_\pi$ and $T \gg \mu_B, m_\pi$. 

Looking first at the $T\ll \mu_B, m_\pi$ limit, we make use of the
asymptotic  series of $K_2$ and find,
\beqn
g_0^Q(T) & = & \sqrt{2} ~ T^4 \left(\frac{m_\pi}{\pi T} \right)^{3/2}
\sum_{n=1}^\infty \frac{1}{n^{5/2}}e^{(2\mu_B-m_\pi)n/T}
\left[1+\frac{15 T}{8 n m_\pi 
} + \ldots \right] \\
& = & \sqrt{2} ~ T^4 \left(\frac{m_\pi}{\pi T} \right)^{3/2}
\left[{\rm Li}_{5/2}(e^{(2\mu_B-m_\pi)/T})
+\frac{15 T}{8 m_\pi} {\rm Li}_{7/2}(e^{(2\mu_B-m_\pi)/T})+ \ldots \right] .
\nn
\eeqn
The polylogarithm in the second line is defined by,
\be
\label{defLi}
{\rm Li}_n(z) \equiv \sum_{k=1}^\infty \frac{z^k}{k^n} \ .
\ee
For $T\ll m_\pi-2\mu_B$, the sum is dominated by the $n=1$ term. When 
$2\mu_B\to m_\pi$, the difference $m_\pi-2\mu_B$ can become comparable 
to or even smaller than $T$. 
If so, more terms will become important in the sum over $n$. 
This is not the case for the $P$ modes as their masses are independent of
$\mu_B$ when $\alpha=0$. Hence, for all $T\ll m_\pi$ we find,
\begin{eqnarray}
 g_0^P&=&\frac{T^4}{\sqrt{2}} \left( \frac{m_\pi}{\pi T} \right)^{3/2} 
              e^{-m_\pi/T}\left [ 1 + \frac {15 T}{8 m_\pi} + \cdots
\right ] \ . 
\label{gqasym}
\end{eqnarray}

In the limit $T \gg \mu_B$, the integrated propagators are given by,
\begin{eqnarray}
  g_0^P&=&-\frac{T}{\pi^2} \int dp ~ p^2 
           \ln [ 1-\exp(-\sqrt{p^2+m_\pi^2}/T) ] 
       =\frac{\pi^2 T^4}{45}-\frac{m_\pi^2 T^2}{12} +\ldots,  \\
  g_0^Q&=&-\frac{T}{\pi^2} \int dp p^2 \left(
           \ln [ 1-\exp((-2\mu_B-\sqrt{p^2+m_\pi^2})/T) ]  \right. \nonumber \\
    && \left. \hspace{4cm}  + \ln [ 1-\exp((2\mu_B-\sqrt{p^2+m_\pi^2})/T) 
\right) +\ldots
   \nonumber \\
       &=& \frac{2 \pi^2 T^4}{45}-\frac{m_\pi^2 T^2}{6}
          +\frac{4 \mu_B^2 T^2}{3}+\ldots \ .  
\end{eqnarray}
The two terms in $g_0^Q$ give the respective contributions from the diquarks
and the anti-diquarks. 

Having calculated the non-zero temperature contributions to the free
energy in the normal phase we now turn to the 
condensates. The baryon density is defined as,
\be
n_B\equiv -\frac{d \Omega}{d \mu_B} \ .
\ee
In the normal phase only the $Q$-modes depend on $\mu_B$. In the low
temperature limit, $T\ll m_\pi$, we thus find,
\be
n_B=\frac{1}{4}N_f(N_f-1)\left(\frac{2m_\pi T}{\pi}\right)^{3/2}{\rm
  Li}_{3/2}(e^{(2\mu_B-m_\pi)/T}) \ . 
\label{nBnormalNc=2}
\ee
This result agrees with the standard form of the density of a relativistic
Bose gas (see eg. \cite{HWjmp}).

The calculation of the temperature dependence of the chiral condensate 
to 1-loop order is
straightforward as well. Using the definition,
\be
\langle \bar \psi \psi \rangle \equiv -\frac{\d \Omega}{\d m} \ ,
\ee
we find,
\be
\langle \bar \psi \psi \rangle (T) & = &\langle \bar \psi \psi \rangle_0
        +\frac12 (N_f^2-1) e^{-m_\pi/T} \frac{G}{2 F^2 \sqrt{m_\pi}} \left(
\frac{T}{2\pi}\right)^{3/2} (3 T-2 m_\pi)  \\
 &&        +\frac14 N_f(N_f-1) \frac{G}{F^2 \sqrt{m_\pi}}  \left(
\frac{T}{2\pi}\right)^{3/2}  \Bigg(3 T{\rm  Li}_{5/2}(e^{(2 \mu_B-m_\pi)/T})
\nn\\
 && \hspace{4cm}
-2 m_\pi
{\rm Li}_{3/2}(e^{(2 \mu_B-m_\pi)/T}) \Bigg) \ . \nn
\label{psibarpsinormalNc=2}
\ee
The zero temperature part, $\langle{\bar \psi} \psi \rangle_0$, was evaluated 
in \cite{STV1},
\begin{eqnarray}
  \label{qbarq-normalP}
\langle \bar \psi \psi \rangle_0= 2 N_f G
\left (1+2\frac{M^2}{F^2}\left [ 
8 N_f L_6 ^r+2 L_8^r +H_2^r 
-\frac {2N_f^2-N_f-1}{128\pi^2N_f}  \ln \frac{M^2}{\Lambda^2} \right ]
\right ) \ . 
\end{eqnarray}
Since we are studying the limit $T\ll m_\pi$ eq.
(\ref{psibarpsinormalNc=2}) can be simplified to,
\be
\langle \bar \psi \psi \rangle (T) & = &\langle \bar \psi \psi \rangle_0
        -\frac12 (N_f^2-1) e^{-m_\pi/T} \frac{G\sqrt{m_\pi}}{ F^2} \left(
\frac{T}{2\pi}\right)^{3/2}    \\
 &&        -\frac12 N_f(N_f-1) \frac{G\sqrt{m_\pi}}{F^2}  \left(
\frac{T}{2\pi}\right)^{3/2}    
{\rm Li}_{3/2}(e^{(2 \mu_B-m_\pi)/T}) \ . \nn 
\label{psibarpsinormalNc=2Simplified}
\ee
The $T\neq 0$ correction is negative. The $\mu_B$ dependence of the
$T\neq 0$ correction is  through the 
${\rm Li}_{3/2}(e^{(2\mu_B-m_\pi)/T})$ factor of the $Q$ modes. The 
corrections are finite and for $(m_\pi-2\mu_B)/T\sim 1$ the 
$\mu_B$-dependent term dominates the exponentially suppressed 
contribution of the $P$-modes.


\subsection{The Phase with Condensed Diquarks at $T\neq 0$}

In the phase of condensed diquarks, the free energy is found to be
given by \cite{STV1},
\begin{eqnarray}
  \label{FrEnD}
  \Omega&=& -2 N_f F^2 \Big(\frac12(M_1^2+M_2^2)+\frac 14M_3^2
  \Big)\nn \\ 
&&-\frac14 N_f (N_f+1) G_0^{P_S}-\frac14 (N_f(N_f-1)-2)
G_0^{P_A}-\frac14 N_f (N_f-1) G_0^Q  \nonumber\\
&& -2 N_f (M_1^2-M_2^2)^2 \Big(L_0+2 N_f L_1+2 N_f L_2+L_3 \Big) 
\\
&&-4 N_f (M_1^2-M_2^2)(M_2^2 +\frac 14 M_3^2) \Big(2 N_f
L_4+L_5 \Big) \nonumber \\
&& -8 N_f (M_2^2 +\frac14 M_3^2 )^2 \Big(2 N_f L_6+L_8
\Big)-2
N_f \tilde M^4 \Big(-2 L_8+H_2 \Big)  \nn \\
&& + {\cal O}(p^6)  \ . \nn
\end{eqnarray}
{}The one-loop integrals of the inverse propagators (\ref{invPropD})
are given by,
\begin{eqnarray}
  \label{Pmodes}
  G_0^{P_S}&=&-\sumint \frac{d^dp}{(2 \pi)^d}
 \ln \Big( p^2+M_1^2 + \frac{1}{4}M_3^2 \Big)  \ , \\
  G_0^{P_A}&=&-\sumint \frac{d^dp}{(2 \pi)^d}
 \ln \Big( p^2+M_2^2 + \frac{1}{4}M_3^2 \Big) \ ,
\end{eqnarray}
and, 
\begin{eqnarray}
  \label{Qmode}
  G_0^Q&=&-\sumint \frac{d^dp}{(2 \pi)^d}
 \ln \Big( (p^2+M_1^2)
(p^2+M_2^2)      +p_0^2 M_3^2 \Big) \ .
\end{eqnarray}

The decomposition of $G_0^{P_S}$ and $G_0^{P_A}$ into a temperature 
independent part $\Delta_0$ and a temperature dependent part $g_0(T)$ 
is completely analogous to the evaluation of the $P$-modes in the normal 
phase, see (\ref{DelPN}), (\ref{gPN}).
The $Q$-modes, however, do not have the standard 
$p^2+m_\pi^2$ form, and the
separation $G_0^Q(T)=\Delta^Q_0+g^Q_0(T)$ is non-trivial. Fortunately this
technical problem can be dealt with in the physically interesting region
for which $\mu_B \simeq m_\pi/2=\mu_c(T=0)$. In this region the rotation 
angle $\alpha$ introduced in (\ref{oS}) is necessarily small and may be 
used as an expansion parameter. In the following we focus on this regime.

If we define, 
\be
\bar \mu\equiv (\mu_B-m_\pi/2)/m_\pi \ ,
\label{Bmu}
\ee
we know from the
zero-temperature results \cite{KSTVZ,STV1} that
$\bar\mu_B\sim\alpha^2$ and that $\phi\sim\alpha^3$.  We now compute
the temperature-dependent part of 
(\ref{Qmode}) up to sixth order in $\al$. This allows us to make the 
decomposition $G_0^Q(T)=\Delta^Q_0+g^Q_0(T)$ in the region  
$\mu_B \simeq \mu_c(T=0)$.

The expansion of $G_0^Q$ is given by,
\begin{eqnarray}
  G_0^Q&=&-\sumint \frac{d^dp}{(2 \pi)^d}
 \ln \Big[ \Big((p_0+i m_\pi y)^2 +\vec{p}^2+m_\pi^2 z^2 \Big) 
\Big((p_0-i m_\pi y)^2+\vec{p}^2+m_\pi^2 z^2 \Big) \\
&&\hspace{3cm} -\frac14 (M_1^2-M_2^2)^2] +O(\al^8)
\nonumber\\ 
&=&
  \label{Qexp}
-\sumint \frac{d^dp}{(2 \pi)^d}
 \ln \Big[\Big((p_0+i m_\pi y)^2 +\vec{p}^2+m_\pi^2 z^2 \Big) 
\Big((p_0-i m_\pi y)^2 +\vec{p}^2+m_\pi^2 z^2 \Big) \Big] \\
&&+\frac14 (M_1^2-M_2^2)^2 \sumint \frac{d^dp}{(2 \pi)^d}
\frac{1}{\Big((p_0+i m_\pi y)^2 +\vec{p}^2+m_\pi^2 z^2 \Big) 
\Big( (p_0-i m_\pi y)^2+\vec{p}^2+m_\pi^2 z^2\Big)} \nn \\
&&+O(\al^8) \ , \nonumber
\end{eqnarray}
where we have defined,
\begin{eqnarray}
  y&\equiv&\frac12 \frac{M_3}{m_\pi}, \\
  z&\equiv&\frac1m_\pi \sqrt{ \frac12 (M_1^2+M_2^2)+\frac14 M_3^2} \nn \ .
\end{eqnarray}

The first term looks exactly like the  $Q$-propagator at the origin in 
the normal phase (cf.~(\ref{helpfulEq})). We find that the $T$-dependent 
part of the first integral in (\ref{Qexp}) is given by 
\begin{eqnarray}
{\cal I}(T)&\equiv&-\sumint \frac{d^dp}{(2 \pi)^d}
 \ln \Big[ \Big((p_0+i m_\pi y)^2+\vec{p}^2+m_\pi^2 z^2\Big)  
\Big((p_0-i m_\pi y)^2+\vec{p}^2+m_\pi^2 z^2\Big) \Big] \nn \\
&=&\frac{2}{\pi^2} \sum_{n=1}^{\infty} \left(\frac{m_\pi zT}{n}
  \right)^2 \cosh\Big(\frac{m_\pi y n}{T}   \Big) K_2\Big( \frac{m_\pi zn}{T}
  \Big).
\label{calI}
\end{eqnarray}

The $n\neq0$ terms in the second integral in
(\ref{Qexp}) are finite and can be rewritten as, 
\begin{eqnarray}
  {\cal J}(T)
&\equiv&\sumint \frac{d^dp}{(2\pi)^2} \frac{1}{\Big((p_0+i m_\pi
    y)^2 +\vec{p}^2+m_\pi^2 z^2 \Big)  
\Big( (p_0-i m_\pi y)^2+\vec{p}^2+m_\pi^2 z^2\Big)} \nn \\
&=&\frac1{8m_\pi^2} \sumint \frac{d^dp}{(2\pi)^2} \frac1{p_0^2} \left(
  \frac1z   \frac{\partial}{\partial z}+ \frac1y \frac{\partial}{\partial y}
\right) \ln\Big[ \Big((p_0+i m_\pi
    y)^2 +\vec{p}^2+m_\pi^2 z^2 \Big)  \\ 
&& \hspace{7cm}
\Big( (p_0-i m_\pi y)^2+\vec{p}^2+m_\pi^2 z^2\Big) \Big] \nn \\
&=&-\frac1{4 \pi^2 m_\pi^2} \sum_{n=1}^\infty \frac1{(2\pi nT)^2}
 \left( \frac1z
  \frac{\partial}{\partial z}+ \frac1y \frac{\partial}{\partial y}
\right) \left(\left(\frac{m_\pi zT}{n} 
  \right)^2 \cosh\Big(\frac{m_\pi y n}{T}   \Big) K_2\Big( \frac{m_\pi
    zn}{T}   \Big) \right) \nn \\
&=& \frac{m_\pi z}{16 \pi^4 yT} \sum_{n=1}^\infty n^{-3}
\Big[ y \cosh \Big(\frac{m_\pi yn}T\Big) K_1\Big(\frac{m_\pi zn}T\Big)
-z \sinh 
\Big(\frac{m_\pi yn}T\Big) K_2\Big(\frac{m_\pi zn}T\Big) \Big]. \nn
\end{eqnarray}
To go from the second to the third line, we have used the same
steps as in (\ref{gQ3}) and (\ref{calI}). 

To order $\alpha^4$ we find that the temperature-dependent
 contribution
of the $Q-$modes for $\mu_B\sim m_\pi/2$ is given by 
\begin{eqnarray}
\label{q0Q1}
  g_0^Q(T)&=&\frac{2}{\pi^2} \sum_{n=1}^{\infty} \left(\frac{m_\pi
      zT}{n} 
  \right)^2 \cosh\Big(\frac{m_\pi y n}{T}   \Big) K_2\Big( \frac{m_\pi
    zn}{T} 
  \Big)  \\
&&+(M_1^2-M_2^2)^2 \frac{m_\pi z}{64 \pi^4 yT}  \nn \\
&& \hspace{1cm} \sum_{n=1}^\infty n^{-3} 
\Big[ y \cosh \Big(\frac{m_\pi yn}T\Big) K_1\Big(\frac{m_\pi zn}T\Big)
-z \sinh 
\Big(\frac{m_\pi yn}T\Big) K_2\Big(\frac{m_\pi zn}T\Big) \Big] . \nn
\end{eqnarray}
Inserting this expression into $\Omega$ of (\ref{FrEnD}) gives the free 
energy for $\mu_B\simeq \mu_c(T=0)$. The free energy gives an effective 
potential for $\al$ as a function of $T$ and $\mu_B$. Using this effective 
potential we now study the diquark phase transition in the $(\mu_B,T)$-plane.

\section{Second Order Phase Transition and Tricritical Point}

Recent lattice simulations \cite{KTS} of two-color QCD at non-zero 
temperature and baryon chemical potential have shown that the phase 
transition between the normal phase and the superfluid phase is of second
order for small temperatures and chemical potentials close to
$m_\pi/2$ and that this second order phase transition line in the
$(\mu_B,T)$-plane ends in a tricritical point. In this section we show
that Chiral 
Perturbation Theory confirms the existence of this tricritical
point. We shall use the effective potential for the orientation angle $\alpha$.
For the shortcomings and problems associated with the use of 
the effective potential approach to determine the properties of the
phase diagram, we refer  to \cite{Baym}. 

The sum in (\ref{q0Q1}) is not in general analytic but can be evaluated in
the high and low temperature limits. Only the  low temperature limit is
of relevance in this case since the critical temperature, $T_c$, is much
smaller than $m_\pi$ for the values of $\mu_B$ under consideration, 
that is $T_c(\mu_B\sim m_\pi/2)\ll m_\pi$. (This is a well known property of
Bose condensation: $T_c\ll M$ if $n_B\ll M^3$, see eg.~\cite{HWprl}.)
In setting up the low temperature limit there are two choices,
$T(\mu_B\sim m_\pi/2)\ll m_\pi (z-y)\ll m_\pi$ or $m_\pi (z-y)\ll
T\ll m_\pi z$. In order to separate 
them it is useful to know the leading order in the expansion of
$y$ and $z$ and $z-y$ in powers of $\alpha$ and $\bar \mu$. To order
$\alpha^4$ they are given by,
\beqn
\label{z-y}
z & =& 1+\bar\mu \al^2 -\frac{1}{16}\al^4+\frac12 \al \phi +\ldots ,\nn
\\
y & = & 1+2\bar\mu-\frac12\al^2-\bar\mu \al^2+\frac{1}{24}\al^4+\ldots , \\
z-y & =& -2\bar\mu
+\frac12\al^2+\frac12\phi\al+2\bar\mu\al^2-\frac{5}{48}\al^4+\ldots ,
\nn \\
(M_1^2-M_2^2)^2&=& m_\pi^4 \alpha^4 + \ldots .  \nn
\eeqn
The critical chemical potential at $T\neq0$ is the value of $\mu$ for which
the vacuum angle $\al$ becomes non-zero. The angle $\al$ is determined by 
minimizing the free energy with respect to $\alpha$. It is thus essential that
we have not fixed $\al$ at the $T=0$ saddle point (\ref{SadPt}).
The dimensionless quantities we expand in here are $\al$, $\bar\mu\sim\al^2$,
$\phi\sim\al^3$. The resulting free energy corresponds to that of a
Landau-Ginzburg approach with $\al$ being the relevant order parameter.

The critical temperature, $T_c$, is the temperature for which the free
energy  
is minimized by $\al=0$.
{}From (\ref{z-y}) we have that $z-y<0$ for $\al=0$. The critical temperature
thus violates the bound $T_c\ll m_\pi (z-y)$. Consequently we can not consistently
evaluate $T_c$ in this limit. Instead we focus on the physically relevant
limit $m_\pi (z-y)\ll T\ll m_\pi z$.

The inverse propagators of the $P_S$ and the $P_A$ modes have the
standard 
$p^2+m_\pi^2$ form (cf. (\ref{invPropD})), and the functions $g_0^{P_S}$ and
$g_0^{P_A}$  in the $T\ll m_\pi z$ limit are given by
\be
g_0^{P_S} & = & T^4 \frac{1}{\sqrt{2}}\left(\frac{\sqrt{M_1^2+M_3^2/4}}{\pi
    T}\right)^{3/2} e^{-\sqrt{M_1^2+M_3^2/4}/T} \\
g_0^{P_A} & = & T^4 \frac{1}{\sqrt{2}}\left(\frac{\sqrt{M_2^2+M_3^2/4}}{\pi
    T}\right)^{3/2} e^{-\sqrt{M_2^2+M_3^2/4}/T} \ .
\ee
As compared to $g_0^Q$ these contributions are exponentially suppressed
and can be neglected.

\noi
Consider the function $g_0^Q$ given in (\ref{q0Q1}) in the limit $T\ll m_\pi$ 
but $T\gg m_\pi(z-y)$. For $T\ll m_\pi z$, the asymptotic expansion of 
$K_2$ can be used, which results in, 
\begin{eqnarray}
  g_0^Q&=&\sqrt{\frac{ 2m_\pi^3 z^3 T^5}{\pi^3}} \sum_{n=1}^{\infty}
  n^{-5/2}  
\cosh\Big(\frac{m_\pi y n}{T}
\Big)\exp\Big[-\frac{m_\pi nz}T\Big]\Big[1+\frac{15T}{8m_\pi
  zn}+\ldots\Big] \nn \\
&&+ \frac{(M_1^2-M_2^2)^2}{512 y \sqrt{2 \pi^7 z m_\pi T}}
\sum_{n=1}^\infty n^{-9/2}   \exp\Big[-\frac{m_\pi nz}T\Big] 
\Big[ y (3 T+8nzm_\pi) \cosh \Big(\frac{m_\pi yn}T\Big) \nn \\
&& \hspace{5cm}-z (15T+8nzm_\pi) \sinh 
\Big(\frac{m_\pi yn}T\Big) \Big].
\end{eqnarray}
Since $z-y\ll z+y$ and using the definition of the polylogarithm
function (\ref{defLi}) we find  that,
\begin{eqnarray}
  g_0^Q&=&\sqrt{\frac{m_\pi^3 z^3 T^5}{2\pi^3}}\Big[ {\rm
  Li}_{5/2}(e^{-m_\pi(z-y)/T})+\frac{15T}{8m_\pi z}{\rm
  Li}_{7/2}(e^{-m_\pi(z-y)/T})+\ldots\Big]  \nn \\
 && -\frac{(M_1^2-M_2^2)^2}{1024 y \sqrt{2 \pi^7 z m_\pi
     T}} \Big[ 8  m_\pi z (z-y) {\rm
   Li}_{7/2}(e^{-m_\pi(z-y)/T}) \\
&&\hspace{4cm}+ 3 T (5z-y) {\rm
   Li}_{9/2}(e^{-m_\pi(z-y)/T})  \Big]+\ldots . \nn
\end{eqnarray}
Now since $m_\pi(z-y) \ll T$
we can expand  Li$_{k}(e^{-x})$ about the origin, see \cite{BMP}. For 
the leading order term we get that,  
\begin{eqnarray}
  \label{QTexp}
 g_0^Q&=&\sqrt{\frac{m_\pi^3 z^3 T^5}{2\pi^3}} \Bigg[
 \zeta(\frac52)
-\zeta(\frac32) \frac{m_\pi(z-y)}T -\Gamma(-\frac32) \left( \frac{m_\pi(z-y)}T
  \right)^{3/2}   \\
&&\hspace{1cm} + \frac12 \zeta(\frac12) \left( \frac{m_\pi(z-y)}T
\right)^2  \Bigg]
-\frac{3(M_1^2-M_2^2)^2}{256}
\zeta(\frac92) \sqrt{\frac{T}{2 \pi^7 m_\pi z}} +O(\al^6). \nn
\end{eqnarray}

{}From now on, and for simplicity, we set $\phi=0$. We are now ready to
evaluate the free energy, 
\begin{eqnarray}
  \label{freeEnT}
  \Omega & = & - N_f F^2 \Big( {\rm const} +(4 \mu_B^2-m_\pi^2)
  \alpha^2-\frac{1}{12}(16 \mu_B^2-M^2)\al^4+2a_4\frac{m_\pi^4}{F^2}\al^4 \\
 && \hspace{5cm} +\frac1{4 F_\pi^2} (N_f-1) g_0^Q \Big) \ , \nn
\end{eqnarray}
where the $T=0$ part, evaluated in \cite{STV1}, is given by,
\be
\label{NLOzero}
a_4 &=&   -\frac{5}{3}(2N_fL^r_4+L^r_5)  
+   \frac 43(2N_fL^r_6+L^r_8) 
 +(L^r_0+2N_fL^r_1+2N_f L^r_2+L^r_3)    \\
&& -\frac {N_f-1}{512 \pi^2 N_f} +\frac{N_f+1}{384
      \pi^2 N_f}  \ln \frac{M^2}{\Lambda^2} \ .  \nn
\ee

The minimum of the free energy determines the ground state of the theory.
With the free energy (\ref{freeEnT}),
the roots of the equation $\partial \Omega/ \partial \alpha=0$ are
given by $\alpha=0$ and the roots of the following quadratic equation
in $\alpha^2$ 
\begin{eqnarray}
  \label{min}
  c_2 + c_4 \alpha^2 + c_6 \alpha^4=0 \ ,
\end{eqnarray}
where, to leading order, we find that
\begin{eqnarray}
  \label{coeff}
  c_2 &=& -\left(32 \sqrt{2 \pi^3}  F^2 \bar{\mu}  - (N_f-1)
     \zeta(\frac32)  \sqrt{m_\pi T^3}\right)^2 +\cdots ,\\
  c_4 &=&  \frac14 (N_f-1)^2 m_\pi T^2  \left( 8 m_\pi \pi - 
  2 m_\pi \zeta(\frac12) \zeta(\frac32) -3 T \zeta^2 (\frac32)
\right)   \nn \\ 
&&+8 \sqrt{2} F^2 \pi^{3/2} \left(32 \sqrt{2 \pi^3}  F^2 \bar{\mu}  - (N_f-1)
     \zeta(\frac32)  \sqrt{m_\pi T^3} \right)  \\ 
&&\hspace{2cm}
  \Bigg(1-\frac{3 (N_f-1) \sqrt{m_\pi^3
      T}}{256 \sqrt{2 \pi^7} F^2} \Big(16 \pi^2 \zeta(\frac12)
    -\zeta(\frac92) \Big)-\frac{a_4 m_\pi^2}{2 \pi^2 F^2} \Bigg)  \nn \\
&&
+\frac{19}{12} \left(32 \sqrt{2 \pi^3}  F^2 \bar{\mu}  - (N_f-1)
     \zeta(\frac32)  \sqrt{m_\pi T^3} \right)^2 + \cdots, \nn \\  
c_6 &=& -32 \pi^3 F^4 + \cdots \ . \nn
\end{eqnarray}

An effective potential of the form $\Omega=c_0+\frac12 c_2 \alpha^2+
\frac14 c_4 \alpha^4+ \frac16 c_6 \alpha^6$ would lead to the same
equation as (\ref{min}). As it is well known from the literature,
the signs of the coefficients $c_2$ and $c_4$ determine the order of
the phase transition (see e.g. \cite{Huang}). If $c_2=0$ and $c_4>0$,
the phase  transition is second order, and if $c_2=0$ and $c_4=0$,
there is a tricritical point. 

\noi
{}From $c_2=0$, we find that the phase transition occurs at,
\begin{eqnarray}
  \label{mu2nd}
  \bar{\mu}_{\rm sec}=\frac{N_f-1}{32 F_\pi^2} \zeta(\frac32)
  \sqrt{\frac{m_\pi T^3}{2 \pi^3}}
  \ .  
\end{eqnarray}
{}From $c_2=0$ and $c_4=0$, we find a tricritical point at,
\begin{eqnarray}
  \label{tricritical}
  \bar{\mu}_{\rm tri}&=&\frac{(N_f-1) m_\pi^2}{6    \sqrt{3}
    \zeta^2(\frac32) F_\pi^2} 
\left(1-\frac{\zeta(1/2) \zeta(3/2)}{4 \pi}\right)^{3/2} \ , \\
 T_{\rm tri}&=& 2 m_\pi \frac{4 \pi-\zeta(1/2) \zeta(3/2)}{3
\zeta^2(3/2)} \ .
\end{eqnarray}

For $T<T_{\rm tri}$, we therefore find a second order phase transition at,
\begin{eqnarray}
  \mu_{\rm sec}(T)&=&m_\pi \Big( \frac12+ \bar \mu_{\rm sec}(T) \Big)\\
&=& \frac{m_\pi}2+\frac1{32 F_\pi^2}
  (N_f-1) \sqrt{\frac{m_\pi^3 T^3}{2 \pi^3}} \zeta(\frac32) \ .
\label{mu_c}
\end{eqnarray}
Alternatively this equation gives $T_{\rm sec}$ as a function of $\mu_B$.  

It is instructive to compare the results obtained with the predictions of a 
dilute Bose gas in the canonical ensemble. In that approach the baryon
density, $n_B$, is fixed and the interaction is determined by the quartic
terms in the chiral Lagrangian. In order to make the comparison recall
that near the phase transition point at $T=0$ we have to leading order 
in $\mu_B -M/2$ \cite{KSTVZ},
\begin{eqnarray}
n_B=32 N_f F^2 (\mu_B-\frac M2 )=32 N_f F^2 M \bar \mu \ .
\end{eqnarray}
Inserting this relation in (\ref{mu_c}) we find that the
critical temperature and the number density are related by, 
\begin{eqnarray}
  n_B=2 N_f (N_f-1) \zeta(\frac32)  \left( \frac{m_\pi
      T_{\rm sec}(n_B)}{2 \pi} \right)^{3/2}, 
\end{eqnarray}
which exactly corresponds to  the relation found in the semi-classical
approach \cite{LL}. 

Finally, notice that both $\bar{\mu}_{\rm tri}$ and $T_{\rm tri}$ vanish in
the chiral limit. This is in complete agreement with the results
obtained within Chiral Perturbation Theory at zero temperature in the
chiral limit: the phase transition is first order and
$\mu_c=\bar{\mu}_c=0$ \cite{KST,KSTVZ}. 
Notice that the tricritical temperature is of the same order as
$m_\pi$. The approximation that we used in the last stage of the
derivation of (\ref{min}) is not fully justified for such high
temperatures. 
The tricritical temperature given above is therefore approximate. In
order to get a more 
precise answer, one needs to numerically compute the different
integrals involved in the free energy. The most important point here
is that Chiral Perturbation Theory confirms the existence of a
tricritical point.  


\section{Order Parameter and Number Density}

The order parameter of the diquark-condensation phase is the diquark
condensate $\langle \psi \psi \rangle=\langle \bar \psi \psi
\rangle_0 \sin \alpha$, where $\langle \bar \psi \psi
\rangle_0$ is the quark-antiquark condensate at zero temperature
and chemical potential. The value of the order parameter close
to the second order phase transition in the $(\mu_B,T)$-plane
is therefore given by the root of (\ref{min}).
To leading order, we obtain, 
\begin{eqnarray}
  \label{alpha}
  \alpha^2&=&8 (\bar{\mu}-\bar{\mu}_{\rm sec}) +\cdots \nn \\
&=&\frac{8}{m_\pi} \left( \mu_B-\frac{m_\pi}2-\frac{N_f-1}{32 F_\pi^2}
  \zeta(\frac32) \sqrt{\frac{m_\pi^3 T^3}{2 \pi^3}} \right).
\end{eqnarray}

In the diquark condensation phase, the number density is given by, 
\begin{eqnarray}
  \label{nBTmu}
  n_B&=&-\frac{d \Omega}{d \mu_B}=-\frac1{M} \frac{d \Omega}{d \bar \mu}
  =4 N_f M F_\pi^2 (\alpha^2+4 \bar{\mu}_{\rm sec}).
\end{eqnarray}
Both $\al$ and $n_B$ decrease with increasing $T$ at fixed 
$\bar\mu>\bar\mu_c$. 
In the zero temperature limit, we recover the results found in
\cite{STV1} for both the order parameter and the number density.

\section{Other ensembles}

So far we have studied two-color QCD with quarks transforming under the 
fundamental representation of the gauge group. The low-energy chiral
Lagrangian  based on the assumption of the spontaneous breaking of the
$SU(2N_f)$  classical chiral symmetry leaving $Sp(2N_f)$ invariant. In
the classification  given in the introduction this is the universality
class labeled by  $\beta_D=1$. 
There exist two other classes of chiral Lagrangians based on spontaneous 
breaking of chiral symmetry. The symmetry breaking patterns are 
$SU(2N_f) \to SO(2N_f)$ and $SU(N_f)\times SU(N_f) \to SU(N_f)$. 
As discussed in the introduction, the first case ($\beta_D=4$) applies to 
QCD with quarks in the adjoint representation of the gauge group,
whereas the  second case
($\beta_D=2$) corresponds to ordinary QCD with $N_c\geq3$ \cite{Jac}. 
It is straightforward to extend the results obtained above to these two 
classes of theories and we do so in the following two subsections.

\subsection{Quarks in the adjoint representation of the gauge group}
\label{sec:adjQCD}

The chiral Lagrangian
based on the coset $SU(2N_f)/SO(2N_f)$ is similar in almost all respects 
to the chiral Lagrangian for two-color QCD. In both theories some
Goldstone modes correspond to diquark states. The diquarks have baryon
charge 2 and  
consequently there is a phase transition at $\mu_B=m_\pi/2$. From the 
study to leading order in \cite{KSTVZ} we know that the propagators in 
the two theories are identical, and only the degeneracy of the modes are
different. 
Extending the results obtained above to $\beta_D=4$ thus simply amounts 
to changing appropriate combinatorial factors. For the convenience of the
reader we explicitly report the results.

In the $\al=0$ phase the baryon density is (this is the analogue of 
(\ref{nBnormalNc=2})),
\be
n_B\equiv\frac{1}{4}N_f(N_f+1)\left(\frac{2m_\pi T}{\pi}\right)^{3/2}
{\rm Li}_{3/2}(e^{(2\mu_B-m_\pi)/T}) \ .
\label{nBnormalAdj}
\ee
The chiral condensate is ($\al=0$),
\be
\langle \bar \psi \psi \rangle (T) & = &\langle \bar \psi \psi \rangle_0
        +\frac12 (N_f^2-1) e^{-m_\pi/T} \frac{G}{2 F^2 \sqrt{m_\pi}} \left(
\frac{T}{2\pi}\right)^{3/2} (3 T-2 m_\pi)  \\
 &&        +\frac14 N_f(N_f+1) \frac{G}{F^2 \sqrt{m_\pi}}  \left(
\frac{T}{2\pi}\right)^{3/2}  \Bigg(3 T{\rm  Li}_{5/2}(e^{(2 \mu_B-m_\pi)/T})
\nn\\
 && \hspace{7cm}
-2 m_\pi
{\rm Li}_{3/2}(e^{(2 \mu_B-m_\pi)/T}) \Bigg) \nn \ .
\label{psibarpsinormalNc=2a}
\ee

The free energy of the diquark condensation phase is given by,
\begin{eqnarray}
  \label{freeEnT(4)}
  \Omega = - N_f F^2 \Big( {\rm const} +(4 \mu_B^2-m_\pi^2)
  \alpha^2-\frac{1}{12}(16 \mu_B^2-M^2)\al^4
+\frac1{4 F_\pi^2} (N_f+1) g_0^Q \Big) \ , 
\end{eqnarray}
where $g_0^Q$ is the same as in the two-color case and is given in
(\ref{QTexp}). We ignored the zero-temperature contribution
that corresponds to (\ref{NLOzero}) since, as in the two-color case
studied in the previous sections,  it does not contribute to
the critical points to leading order. This free energy is extremely
similar to (\ref{freeEnT}). The analysis of the phase diagram
proceeds in the same way as in Section~6.

We find that the critical chemical potential for the second order phase
transition is given by,
\begin{eqnarray}
  \mu_{\rm sec}(T)
= \frac{m_\pi}2+\frac{N_f+1}{32 F_\pi^2}
  \sqrt{\frac{m_\pi^3 T^3}{2 \pi^3}} \zeta(\frac32) \ ,
\label{mu_cAdj}
\end{eqnarray}
and that the phase transition turns first order at the tricritical
point given by,
\begin{eqnarray}
  \label{tricriticalbeta4}
  \mu_{\rm tri}&=&\frac{m_\pi}2 + \frac{(N_f+1) m_\pi^3}
{6    \sqrt{3}  \zeta^2(\frac32) F_\pi^2}
\left(1-\frac{\zeta(1/2) \zeta(3/2)}{4 \pi}\right)^{3/2} \ , \\
 T_{\rm tri}&=& 2 m_\pi \frac{4 \pi-\zeta(1/2) \zeta(3/2)}{3
\zeta^2(3/2)} \ .
\end{eqnarray}

The order parameter close to the second order phase transition is
readily obtained in the same way as in the two-color case. 
To leading order, we obtain,
\begin{eqnarray}
  \label{alpha2}
  \alpha^2
=8 (\bar{\mu}-\bar{\mu}_{\rm sec}),
\end{eqnarray}
where $\bar{\mu}_{\rm sec}$ is given by (\ref{mu_cAdj}), (\ref{Bmu}).
In the diquark condensation phase, the number density is given by, 
\begin{eqnarray}
  \label{nBTmu1}
  n_B&=&
  4 N_f M F_\pi^2 (\alpha^2+4 \bar{\mu}_{\rm sec}).
\end{eqnarray}

\subsection{Ordinary QCD with $\mu_I\neq0$}
\label{sec:QCDmuI}

In QCD with $N_c\geq3$ and quarks in the fundamental representation of 
the gauge group the Goldstone manifold is $SU(N_f)$. The diquarks are 
no longer Goldstone modes and the chiral Lagrangian does not couple 
to $\mu_B$. The pions do however have non-trivial isospin and consequently 
the chiral Lagrangian couples to a chemical potential,
$\mu_I=(\mu_u-\mu_d)/2$,  
for the third component of isospin. This theory is equivalent to
QCD at nonzero baryon chemical potential but with a fermion
determinant replaced by its absolute value.
It was shown in \cite{TV,Misha-Son} 
that pion condensation occurs for $\mu_I\geq m_\pi$ at $T=0$. An isospin 
chemical potential also has non-trivial effects for $\beta_D=1$ and 
$\beta_D=4$. As was shown in \cite{SSS} the competition between $\mu_I$ 
and $\mu_B$ can lead to novel first order phase transitions. Similar 
phenomena occur in QCD with $N_c=3$ and $N_f=3$ \cite{Dominique}. 
In this system the competition is established 
between $\mu_I$ and the strangeness chemical potential $\mu_S$. 
Here we shall give the leading temperature effects for QCD with $N_c\geq3$ 
and quarks in the fundamental representation with a generalized isospin 
chemical potential.
For an an even number of flavors $N_f=2n$ we take $n$ flavors with chemical 
potential $\mu_I$ and $n$ flavors with chemical potential $-\mu_I$.\\

The particular interest in this system comes from that fact that the fermion
determinant in the path integral of the Euclidean partition function is 
positive, see \cite{AKW}. The first lattice study of QCD at non-zero $\mu_I$ 
and $T$ appeared recently \cite{KS2}. The similarity with two color 
QCD at $\mu_B\neq0$ is striking. Again the chiral Lagrangian is completely
analogous to QCD with two colors and describes the same physics.
At $T=0$ and $\mu_I=m_\pi$ there is a 
second order phase transition to a pion condensed  phase and with
increasing temperature this phase transition turns first order. \\

At $\mu_I=0$ and $m=0$ this system has an  $SU(N_f)\times SU(N_f)\times
U(1)$  classical  symmetry which is assumed to be spontaneously broken to 
$SU(N_f)\times U(1)$ by the vacuum. There are $N_f^2-1\equiv 4n^2-1$ 
Goldstone modes which organize themselves into $n^2$ isotriplets and 
$n^2-1$ isosinglets. At $\mu_I\neq0$, the $SU(N_f)\times U(1)$ is 
explicitly broken down to $SU(n)\times SU(n)\times U(1)$ leaving only 
internal rotations in the $u$ and the $d$ sector invariant. At 
$\mu_I=m_\pi$ the $SU(n)\times SU(n)\times U(1)$ invariance is 
spontaneously broken down to $SU(n)$ leaving $n^2$ massless modes. 
The leading order chiral Lagrangian is \cite{TV,Misha-Son},
\be
{\cal L}^{(2)} = \frac{F^2}{4}\Tr\nabla_\nu \Sigma \nabla_\nu\Sigma^\dagger
-\frac{F^2 M^2}{2}{\rm Re}\Tr\Sigma \ , 
\ee
where,
\be
\nabla_\nu \Sigma = \d_\nu \Sigma
-\delta_{\mu0}\frac{\mu_I}{2}(\tau_3\Sigma-\Sigma\tau_3) \ .
\ee
The isospin charge matrix is denoted $\tau_3$.
Since the isospin charge of the pions is equal to one we redefine
$\bar\mu$ as, 
\be
\label{BmuI}
\bar \mu\equiv (\mu_I-m_\pi)/m_\pi . 
\ee
We  obtain the free energy for $T<<M$, 
\be
\Omega&=& - N_f F^2 \Big(\frac12(M_1^2+M_2^2)+\frac 14M_3^2
  \Big) -\frac12 n G_0^Q,
\ee
where, as in \cite{SSS},
\begin{eqnarray}
\label{massQI}
M_1^2&\equiv&\tilde M^2 \cos (\alpha-\phi)-\mu_I^2 \cos 2\alpha \ , \nonumber
\\ 
M_2^2&\equiv &\tilde M^2 \cos (\alpha-\phi)-\mu_I^2 \cos^2\alpha \ , \\
M_3^2&\equiv &4 \mu_I^2 \cos^2 \alpha \ . \nonumber
\end{eqnarray}

In the $\al=0$ phase the isospin density is, 
\be
n_I\equiv\frac{1}{8}n^2\left(\frac{2m_\pi T}{\pi}\right)^{3/2}{\rm
  Li}_{3/2}(e^{(\mu_I-m_\pi)/T}) \ . 
\label{nInormal}
\ee
The chiral condensate is ($\al=0$),
\be
\langle \bar \psi \psi \rangle (T) & = &\langle \bar \psi \psi \rangle_0
        +\frac12 (2n^2-1) e^{-m_\pi/T} \frac{G}{2 F^2 \sqrt{m_\pi}} \left(
\frac{T}{2\pi}\right)^{3/2} (3 T-2 M)  \\
 &&        +\frac12 n^2 \frac{G}{F^2 \sqrt{m_\pi}}  \left(
\frac{T}{2\pi}\right)^{3/2}  (3 T{\rm  Li}_{5/2}(e^{(\mu_I-m_\pi)/T})-2 m_\pi
{\rm Li}_{3/2}(e^{(\mu_I-m_\pi)/T})) \nn \ .
\label{psibarpsinormalmuI}
\ee

The free energy of the diquark condensation phase is given by,
\begin{eqnarray}
  \label{freeEnT(2)}
  \Omega = -\frac{n}2 F^2 \Big( {\rm const} +(\mu_I^2-m_\pi^2)
  \alpha^2-\frac{1}{12}(4
  \mu_I^2-M^2)\al^4
+\frac{n}{F_\pi^2} g_0^Q \Big) \ , 
\end{eqnarray}
where $g_0^Q$ is the same as in the two-color case and is given in
(\ref{QTexp}), but with, 
\begin{eqnarray}
z & =& 1+\frac12 \bar\mu \al^2 -\frac{1}{16}\al^4+\ldots, \nn
\\
y & = & 1+\bar\mu-\frac12\al^2-\frac12 \bar\mu
\al^2+\frac{1}{24}\al^4+\ldots,  \\
M_1^2-M_2^2&=& m_\pi^2 \alpha^2+\ldots , \nn
\end{eqnarray}
instead of (\ref{z-y}) and $\bar{\mu}$ as defined in (\ref{BmuI}).
We have ignored the zero-temperature contribution
that corresponds to (\ref{NLOzero}) since it does not contribute to
the critical points to leading order as found in Section~6. This free
energy is extremely 
similar to (\ref{freeEnT}). The analysis of the phase diagrams
proceeds in the same way as in Section~6.

We find that the critical isospin chemical potential for the line of
second order  phase transitions is given by,
\begin{eqnarray}
  \mu_{\rm sec}(T)&=& m_\pi+\frac n{4 F_\pi^2}
  \sqrt{\frac{m_\pi^3 T^3}{2 \pi^3}} \zeta(\frac32) \ ,
\label{muIc}
\end{eqnarray}
and that this second order phase transition line ends in a tricritical
point at
\begin{eqnarray}
  \label{tricriticalbeta2}
  \mu_{\rm tri}&=& m_\pi+4\frac{n m_\pi^3}{3    \sqrt{3}
    \zeta^2(\frac32) F_\pi^2} 
\left(1-\frac{\zeta(1/2) \zeta(3/2)}{4 \pi}\right)^{3/2} \ , \\
 T_{\rm tri}&=& 2 m_\pi \frac{4 \pi-\zeta(1/2) \zeta(3/2)}{3
\zeta^2(3/2)} \ .
\end{eqnarray}

The order parameter close to the second order phase transition is
easily obtained in the same way as in the two-color case. 
To leading order, we obtain,
\begin{eqnarray}
  \label{alpha3}
  \alpha^2
=4 (\bar{\mu}-\bar{\mu}_{\rm sec}),
\end{eqnarray}
where $\bar{\mu}_{\rm sec}$ is given by (\ref{muIc}), (\ref{BmuI}).
In the superfluid phase the number density is given by, 
\begin{eqnarray}
  \label{nBTmu2}
  n_I
  =n M F_\pi^2 (\alpha^2+2 \bar{\mu}_{\rm sec}).
\end{eqnarray}

\section{Summary and Conclusions}

In this article we have studied the phase diagram of various QCD-like
theories at non-zero chemical potential and non-zero temperature.  
These studies have been performed by means of
chiral Lagrangians with the symmetries of the
microscopic theories. The chiral Lagrangians describe the low-energy dynamics
of the Goldstone modes and
are limited to phases where the symmetry of the
microscopic Lagrangian is spontaneously broken by the ground state. 
Although such theories cannot be used to study
the high temperature restoration of the broken symmetry, they are very
powerful to study the different low-energy phases. 

Using Chiral Perturbation Theory, we have demonstrated that the phase
diagram of QCD with 
two colors and quarks in the fundamental representation at non-zero
temperature and baryon chemical potential has a very rich
structure. At low temperature, there are two different phases: a
``normal'' phase with a non-zero quark-antiquark condensate, and a
superfluid diquark condensation phase with a non-zero diquark
condensate. For 
small enough temperatures, we have found that a second order phase
transition separates these two phases. Our result for this second order
phase transition line in the $(\mu_B,T)$-plane are in complete
agreement with the semi-classical dilute gas approximation. However,
we have shown analytically that 
this second order phase transition line ends in a tricritical
point when the chemical potential is increased. Above this point,
the two phases are
separated by a first order phase transition. All our results are
solely based on symmetry and are therefore very robust. They are in
complete agreement with recent lattice simulations \cite{KST,KST2} of 
this theory at non-zero temperature and chemical potential and with recent 
Random Matrix \cite{KTV} studies.

We have also analyzed the phase diagram of theories that share many
properties of QCD with two
colors and quarks in the fundamental representation at non-zero
temperature and baryon chemical potential. We have shown that the
phase diagram of QCD with any number of colors and quarks in the
adjoint representation at non-zero temperature and baryon chemical
potential, as well as the
phase diagram of QCD with three colors and quarks in the
fundamental representation at non-zero temperature and isospin chemical
potential are very similar to the phase diagram of QCD with two
colors and quarks in the fundamental representation at non-zero
temperature and baryon chemical potential that was described above. We
again found a second 
order phase transition separating the normal phase from a superfluid
phase at small temperature that ends in a tricritical point when the
temperature is increased. 

These different QCD-like theories all contain
Goldstone modes in their spectrum at zero temperature and zero
chemical potential. They all  react similarly upon the
introduction of a chemical potential associated with a quantum number 
carried by at least one of these Goldstone modes: At low enough
temperature and high enough
chemical potential, the ground sate of the theory is a superfluid. The
normal and superfluid phases are separated by a second order phase
transition at small temperatures. This second order phase transition
lines ends in a tricritical point and the phase transition is first
order when the temperature increases.
This remains true for any physical system with spontaneous breaking
of a global symmetry, and their phase diagram at non-zero temperature
and chemical potential will be similar to the one described above.
In all these theories, the
order parameter of the superfluid phase and the number density also
behave similarly. These results are again in complete
agreement with available lattice simulations.
The results of this article could be used as non-trivial tests for
new simulation schemes developed to solve the sign problem
\cite{FodorKatz}.

{\bf Acknowledgments:} 
G. Baym, B. Klein, J. Kogut, J.T. Lenaghan, R. Pisarski,
B. Vanderheyden,  are acknowledged 
for useful discussions. D.T. is supported in part by
``Holderbank''-Stiftung. This work was
partially supported by the US DOE grant DE-FG-88ER40388 and by the NSF
grant NSF-PHY-0102409.

\end{document}